\documentclass[prd,amsmath,amssymb,superscriptaddress,preprintnumbers,twocolumn,10pt]{revtex4-1}%superscriptaddress,showpacs,
\usepackage{graphicx}
\usepackage{dcolumn}
\usepackage{bm}
\usepackage{amssymb}
\usepackage{latexsym}
\usepackage{booktabs}
\usepackage{amsmath}
\usepackage{multirow}
\usepackage{url}
\usepackage{footnote}
\usepackage{float}
\usepackage[colorlinks=true, linkcolor=red, citecolor=blue]{hyperref}

\usepackage[normalem]{ulem}
\usepackage{color}
\usepackage{array}
\usepackage{enumerate}
\usepackage{textcomp}
\usepackage{subfigure}

\newcommand{\orcid}[1]{%
  \href{https://orcid.org/#1}{%
    \includegraphics[height=2ex, keepaspectratio]{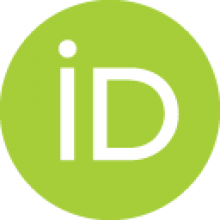}%
  }%
}

\usepackage{natbib}                    
\begin{document}

\title{Factorized neural posterior estimation for rapid and reliable inference of parameterized post-Einsteinian deviation parameters in gravitational waves}

\author{Yong-Xin Zhang\orcid{0009-0001-5942-2991}\email{2400200@stu.neu.edu.cn}}
\affiliation{Liaoning Key Laboratory of Cosmology and Astrophysics, \ College of Sciences, Northeastern University, Shenyang 110819, China}
\author{Tian-Yang Sun\orcid{0009-0002-5109-6420}\email{suntianyang@stumail.neu.edu.cn}}
\affiliation{Liaoning Key Laboratory of Cosmology and Astrophysics, \ College of Sciences, Northeastern University, Shenyang 110819, China}
\author{Chun-Yu Xiong\orcid{0009-0007-7041-9543}}
\affiliation{Liaoning Key Laboratory of Cosmology and Astrophysics, \ College of Sciences, Northeastern University, Shenyang 110819, China}
\author{Song-Tao Liu\orcid{0000-0001-5966-2461}\email{songtaoliu_phy@163.com}}
\affiliation{Liaoning Key Laboratory of Cosmology and Astrophysics, \ College of Sciences, Northeastern University, Shenyang 110819, China}
\author{Yu-Xin Wang\orcid{0000-0002-9592-9087}\email{2210028@stu.neu.edu.cn}}
\affiliation{Liaoning Key Laboratory of Cosmology and Astrophysics, \ College of Sciences, Northeastern University, Shenyang 110819, China}

\author{Shang-Jie Jin\orcid{0000-0003-3697-3501}}
\affiliation{Liaoning Key Laboratory of Cosmology and Astrophysics, \ College of Sciences, Northeastern University, Shenyang 110819, China}

\affiliation{Department of Physics, University of Western Australia, \ Perth WA 6009, Australia}

\author{Jing-Fei Zhang\orcid{0000-0002-3512-2804}}
\affiliation{Liaoning Key Laboratory of Cosmology and Astrophysics, 
\ College of Sciences, Northeastern University, Shenyang 110819, China}
\author{Xin Zhang\orcid{0000-0002-6029-1933}}\thanks{Corresponding author.\\zhangxin@mail.neu.edu.cn}
\affiliation{Liaoning Key Laboratory of Cosmology and Astrophysics, \ College of Sciences, Northeastern University, Shenyang 110819, China}

\affiliation{MOE Key Laboratory of Data Analytics and Optimization for Smart Industry, \ Northeastern University, Shenyang 110819, China}

\affiliation{National Frontiers Science Center for Industrial Intelligence and Systems Optimization, \ Northeastern University, Shenyang 110819, China}

\begin{abstract}
The direct detection of gravitational waves (GWs) by LIGO has confirmed general relativity (GR), but testing GR via GWs requires estimating parameterized post-Einsteinian (ppE) deviation parameters in waveform models. Traditional Bayesian inference methods like Markov chain Monte Carlo (MCMC) provide reliable estimates but suffer from prohibitive computational costs, failing to meet the real-time demands and surging data volume of future GW detectors. Here, we propose a factorized neural posterior estimation framework: we construct independent normalizing flow models for each of the nine ppE deviation parameters, and effectively integrate prior information from other source parameters via a conditional embedding network. Leveraging a hybrid neural network with a convolutional neural network and a Residual Neural Network for feature extraction, our method performs rapid and statistically reliable posterior inference directly from binary black hole signals. Compared to conventional MCMC, our approach achieves millisecond-scale inference time with a speedup factor of $9 \times 10^4$. Comprehensive validations show that the posterior estimates pass the Kolmogorov-Smirnov test and achieve empirical coverage probabilities close to theoretical targets. We apply this method to the GW150914 event and find that the resulting posterior distributions are in good agreement with existing results. This work demonstrates the great potential of deep learning for GW parameter estimation and provides a viable technical solution for real-time GR tests with next-generation detectors.

\end{abstract}
%\pacs{95.36.+x, 98.80.Es, 98.80.-k}
\maketitle
\section{INTRODUCTION}
The direct detection of gravitational waves (GWs) marks a new window for humanity to understand the universe. Following the first direct detection of GWs from the GW150914 binary black hole (BBH) merger by LIGO in 2015 \cite{PhysRevLett.116.061102}, the field has rapidly evolved from the initial goal of detection to an era characterized by the growing catalog of events and precision measurement. The international network comprising LIGO, Virgo, and KAGRA detectors has now detected numerous GW events, with sources including BBH, binary neutron stars (BNS), and black hole-neutron star mergers \cite{LIGOScientific:2025slb,LIGOScientific:2018mvr,KAGRA:2021vkt,LIGOScientific:2020ibl,LIGOScientific:2021usb}. These discoveries not only provide strong confirmation of Einstein's general relativity (GR) but also provide irreplaceable new insights for astrophysics \cite{LIGOScientific:2018cki,Annala:2017llu,Margalit:2017dij,LIGOScientific:2016vpg}, fundamental physics \cite{LIGOScientific:2018dkp,LIGOScientific:2020tif,LIGOScientific:2021sio,Isi:2019aib,Yunes:2016jcc, LIGOScientific:2025rid}, and cosmology \cite{LIGOScientific:2017adf,Wang:2018lun,LIGOScientific:2021aug,Zhang:2019loq,LIGOScientific:2019zcs,Song:2022siz,Zhang:2019ylr,DES:2019ccw,Zhang_2019,LIGOScientific:2018gmd,Li:2023gtu,LIGOScientific:2025jau,Zhang:2018byx,Song:2025bio,Song:2025ddm,Pierra:2026ffj,Pierra:2025hoc,Mastrogiovanni:2023emh,Gray:2023wgj,Li:2025dwz,Li:2025owk} research. GW astronomy is still in its early stages, with observational data and detection capabilities remaining limited. Therefore, further enhancing GW observational capabilities is of critical importance for understanding the origins and distributions of astrophysical compact binary coalescence sources \cite{Mandel:2021smh,vanSon:2021zpk,Broekgaarden:2021efa,Ezquiaga:2022zkx}, measuring cosmological parameters with high precision \cite{Verde:2019ivm,Cai:2016sby,Cai:2017plb,Cai:2017cbj,Cai:2017aea,Jin:2020hmc,DES:2020nay,Jin:2021pcv,Wang:2021srv,Hou:2022rvk,guo2022standard,Jin:2022tdf,Palmese:2021mjm,Jin:2023sfc,Han:2023exn,Jin:2023tou,Schutz:1986gp,Holz:2005df,Nissanke:2009kt,Zhao:2010sz,Zhao:2019gyk,Wang:2019tto,Wu:2022dgy,Wang:2022oou,Jin_2026}, and testing GR \cite{Berti:2015itd,LIGOScientific:2018dkp,LIGOScientific:2019fpa,LIGOScientific:2020tif,LIGOScientific:2021sio,Gong:2021jgg,Gong:2023ffb}. 

As the prevailing theory of gravity, GR has passed extensive experimental tests \cite{Will2014}. GW signals also provide crucial tests for GR under strong-field and high-energy conditions \cite{PhysRevLett.116.061102}. Any potential deviations from GR may manifest as subtle deviations in waveform phase or amplitude. By systematically measuring these deviations, we can probe physics beyond standard gravitational theories. The mainstream verification method involves introducing GR deviation parameters into the GW. These parameters are defined at each Post-Newtonian (PN) order, and determining from the data whether these parameters significantly deviate from the GR prediction \cite{Li_2012, Agathos_2014, Mehta_2023}. Under the parameterized post-Einsteinian (ppE) formalism \cite{Yunes_2009, Cornish_2011, Chatziioannou_2012, Sampson_2013, Yunes_2016, Tahura_2018, Zhang:2019iim}, such deviations are encoded as waveform amplitude and phase corrections, thereby constructing small deviations from GR. Additionally, current research efforts are conducting a broad and multifaceted test of GR using GW data through multiple complementary approaches. Including the generic parametrized modifications, the inspiral–merger–ringdown (IMR) consistency test, spin-induced quadrupole measurements \cite{PhysRevD.103.122002}, and parameterized tests of GW propagation \cite{Zhu:2023rrx}. Although the ppE framework \cite{Yunes_2009, Cornish_2011, Chatziioannou_2012, Sampson_2013, Yunes_2016, Tahura_2018, Zhang:2019iim} can characterize deviations across the entire IMR process, traditional Bayesian analyses (e.g., {\tt LALInference} \cite{PhysRevD.91.042003} and {\tt Bilby} \cite{Ashton_2019}) introduce bias coefficients, expanding the parameter space. For high SNR signals, the number of likelihood estimates required to decompose narrow posteriors increases accordingly \cite{kumar2025acceleratingparameterestimationparameterized}. Researchers have proposed several acceleration strategies \cite{Canizares_2013, Smith_2016, Cornish_2021,zackay2018relativebinningfastlikelihood, Finstad_2020, Morisaki_2021, Leslie:2021ssu, PhysRevD.106.123015,islam2022factorizedparameterestimationrealtime,wong2023fastgravitationalwaveparameter,narola2023relativebinningcompletegravitationalwave}, but their application in GR testing is limited \cite{Adhikari_2022}.

Deep learning has garnered significant attention due to its successful application in GW detection \cite{Sun__2024, PhysRevD.95.104059, PhysRevLett.120.141103, Fan_2019, Krastev_2020, Wang_2020a,
Krastev_2021,Cabero_2020,Wei_2021,Sch_fer_2022,Verma_2022,moreno2021sourceagnosticgravitationalwavedetectionrecurrent,Qiu_2023,PhysRevD.106.122002,Nousi_2023,Ma_2022,Schafer:2021fea,Wang__2024,yun2023detectionextractionparameterestimation, PhysRevD.107.063029,Trovato_2024,stergioulas2024machinelearningapplicationsgravitational, McLeod_2025,wang2024waveformertransformerbaseddenoisingmethod, George_2018, Xia_2021, Jadhav_2021, Zhao_2023, Beveridge_2025, Koloniari_2025, sun2025conditionalvariationalautoencoderscosmological,Wang_2025,Sun_2025,Xiong_2025,Chua_2019,Chua_2020,chen2020machinelearningnanohertzgravitational,Huerta_2021,gunny2021hardwareacceleratedinferencerealtimegravitationalwave,Dax_2023,Langendorff_2023,Boccelli_2023,Koksbang_2023,Lucie_Smith_2024,shih2023fastparameterinferencepulsar,Ocampo_2025,Speri:2022kaq,ribli2018improvedcosmologicalparameterinference,hahn2023rmsscriptsizeimbigcosmological,harvey2024deeplearningalgorithmdisentangleselfinteracting,Liu:2026nhu,Ma:2026qgq,Ma:2025nex}. To overcome the computational limitations of traditional methods, researchers have begun exploring alternative approaches using deep learning to address these challenges. Early attempts primarily framed parameter estimation as classification or regression problems in supervised learning, utilizing neural network architectures such as a convolutional neural network (CNN) to directly predict optimal parameter values from data \cite{PhysRevLett.120.141103}. Although these methods achieve millisecond-scale inference speeds, they typically yield only point estimates. They cannot provide the complete posterior distributions and confidence intervals required for GW data analysis, and are thus inadequate for drawing rigorous scientific conclusions. Variational autoencoders (VAE) have demonstrated their potential for neural networks to learn complete posterior distributions \cite{Gabbard_2021}. Building upon this foundation, the Neural Posterior Estimation (NPE) framework \cite{
Papamakarios:2016ctj,lueckmann2017flexiblestatisticalinferencemechanistic,greenberg2019automaticposteriortransformationlikelihoodfree,Liang:2025yjw}, constructed via normalizing flow \cite{Cranmer_2020,2021Normalizing}, has been proposed as a more general simulation-based inference framework. The DINGO project serves as an exemplary representation of this framework. It has significantly accelerated parameter estimation for signals such as BBH mergers \cite{Dax_2021,Dax:2022pxd} and has further achieved real-time, high-precision full parameter inference for BNS mergers \cite{Dax_2025}. The NPE framework has demonstrated significant potential in simplifying GW model analysis \cite{green2020completeparameterinferencegw150914}. By leveraging deep learning VAE \cite{kingma2022autoencodingvariationalbayes}, the model constructs a continuous latent space mapping the dephasing of several discrete PN models. Furthermore, it enhances the detection of PN deviations with higher PN order corrections and enables more efficient parameter estimation schemes compared to previous implementations \cite{xie2025neuralposteinsteiniantestgeneral}. A key advantage of the NPE framework is that it obviates explicit likelihood functions by learning a direct mapping from simulated data to parameter space, thereby achieving rapid inference \cite{papamakarios2018fastepsilonfreeinferencesimulation,qin2025parameterinferencemicrolensedgravitational}. 

Based on the findings of Refs. \cite{Dax_2021,Dax:2022pxd}, our work proposes a factorizable NPE framework. We design and train a lightweight, dedicated NPE model independently for each ppE parameter. Building upon Residual Neural Network (ResNet) \cite{he2015deepresiduallearningimage}, we employ a hybrid CNN and ResNet front-end network to extract features from observational data for each model. All other parameters are incorporated as conditional information. This design fundamentally circumvents the complex couplings inherent in high-dimensional joint estimation, alleviating the optimization difficulties and mode collapse problems commonly encountered during training. Consequently, this approach significantly enhances the sensitivity to subtle deviations from GR. To evaluate the statistical calibration of the posterior estimates, we performed rigorous Monte Carlo coverage tests \cite{hermans2022trustcrisissimulationbasedinference} for each independent model, specifically by applying the Kolmogorov-Smirnov (KS) test to the test set \cite{Lopes2011}. Compared with methods that accelerate Bayesian sampling using machine learning tools \cite{green2020completeparameterinferencegw150914, Dax_2021, Dax_2025, Garc_a_Quir_s_2025, negri2025neurallikelihoodestimatorsflexible, Chua_2020, Polanska:2024zpn, Andres-Carcasona:2023rnk}, the core innovation of our approach is the implementation of fully amortized NPE. This method uses normalizing flows to directly learn the nonlinear mapping from data to the posterior distribution. It completely circumvents the need for reduced-order model compression, iterative Markov Chain Monte Carlo (MCMC) sampling, or assumptions about parametric distribution forms. An end-to-end optimized deep embedding network automatically extracts features. Its generative capability, when combined with normalizing flows, outputs arbitrarily complex posteriors at millisecond speeds. Our framework yields significant improvements in inference speed, posterior fidelity, and model flexibility.

The subsequent sections are structured as follows. In Section~\ref{sec2}, the data generation process, methodological architecture, and training strategies are described in detail. The analysis of posterior quality, coverage probability, runtime performance, and testing for GW150914 is presented in Section~\ref{sec3}. Finally, potential future research directions are discussed in Section~\ref{sec4}.
\section{METHODOLOGY}\label{sec2}
\subsection{Dataset}\label{sec2.1}
To cover the distribution of genuine GW signals, we constructed a large-scale simulated dataset within a physically plausible parameter space. The parameter space includes 15 fundamental physical parameters \cite{Romero-Shaw:2020owr}  and 9 ppE parameters ($\delta \chi_0$ to $\delta \chi_7$). 
For each ppE parameter, its prior range is chosen to ensure that the contribution of the ppE phase correction remains smaller than the corresponding GR leading term across the entire observational frequency band. Following this principle and taking into account the mass range and frequency window adopted in our simulations, we set uniform priors for the ppE parameters at different PN orders \cite{Chandramouli:2024vhw, Liu:2024xxn}. The fundamental physical parameters include the intrinsic parameters, namely the component masses \(m_1\) and \(m_2\) and the spin angular momenta \(\vec{S}_1\) and \(\vec{S}_2\) of the binary, as well as the extrinsic parameters, which consist of the luminosity distance \(D_L\), the sky location (right ascension \(\alpha\) and declination \(\delta\)), the inclination angle \(\iota\) of the orbital plane with respect to the observer’s line of sight, the polarization angle \(\psi\), and the reference time \(t_c\) and orbital phase \(\phi_c\). The prior distributions of various GR parameters (intrinsic and extrinsic) are the same as in Ref. \cite{LIGOScientific:2020ibl}. In particular, the prior range for the luminosity distance is [100, 6000] Mpc, which naturally generates simulated signals covering a wide range of signal-to-noise ratio (SNR) \cite{Dax_2021}.

Accounting for precession, GW templates were approximated using {\tt PyCBC} \cite{Pycbc_2023} with IMRPhenomXPHM \cite{Pratten:2020ceb}. This is represented in the frequency domain, with $f_{\min} = 20\,\mathrm{Hz}$, $f_{\max} = 1024\,\mathrm{Hz}$ \cite{Dax_2021}. The simulation ultimately produced $1\,\mathrm{s}$ BBH signals with a sampling rate of 4096 Hz. The training set comprises \(2 \times 10^7\) independent samples, generated in parallel across 96 processes for efficiency. Each sample consists of a $1\,\mathrm{s}$ strain time series and its associated parameter $\theta$. This work simulates the observational configuration of the two-detector LIGO network (Hanford and Livingston).
The {\tt aLIGOZeroDetHighPower} model is used as the power spectral density model to simulate noise \cite{LIGO-T0900288}. Based on the specified power spectral density, we employ the noise generation module of {\tt PyCBC} to generate colored noise in the time domain. This process utilizes a random seed to ensure reproducibility. Concurrently, we implement a frequency-domain whitening scheme
\begin{equation}
\tilde{x}(f) = \frac{\tilde{h}(f) + \tilde{n}(f)}{\sqrt{S_n(f)}}.
\end{equation}
Here, $\tilde{h}(f)$ denotes the signal spectrum, $\tilde{n}(f)$ denotes the noise spectrum, and $\tilde{S_n}(f)$ denotes the power spectral density after interpolation. After whitening in the frequency domain, the signal is converted back to the time domain via an inverse Fourier transform, effectively suppressing the dominant influence of low-frequency noise.

\begin{figure*}[!htp]
\includegraphics[width=1.0\textwidth]{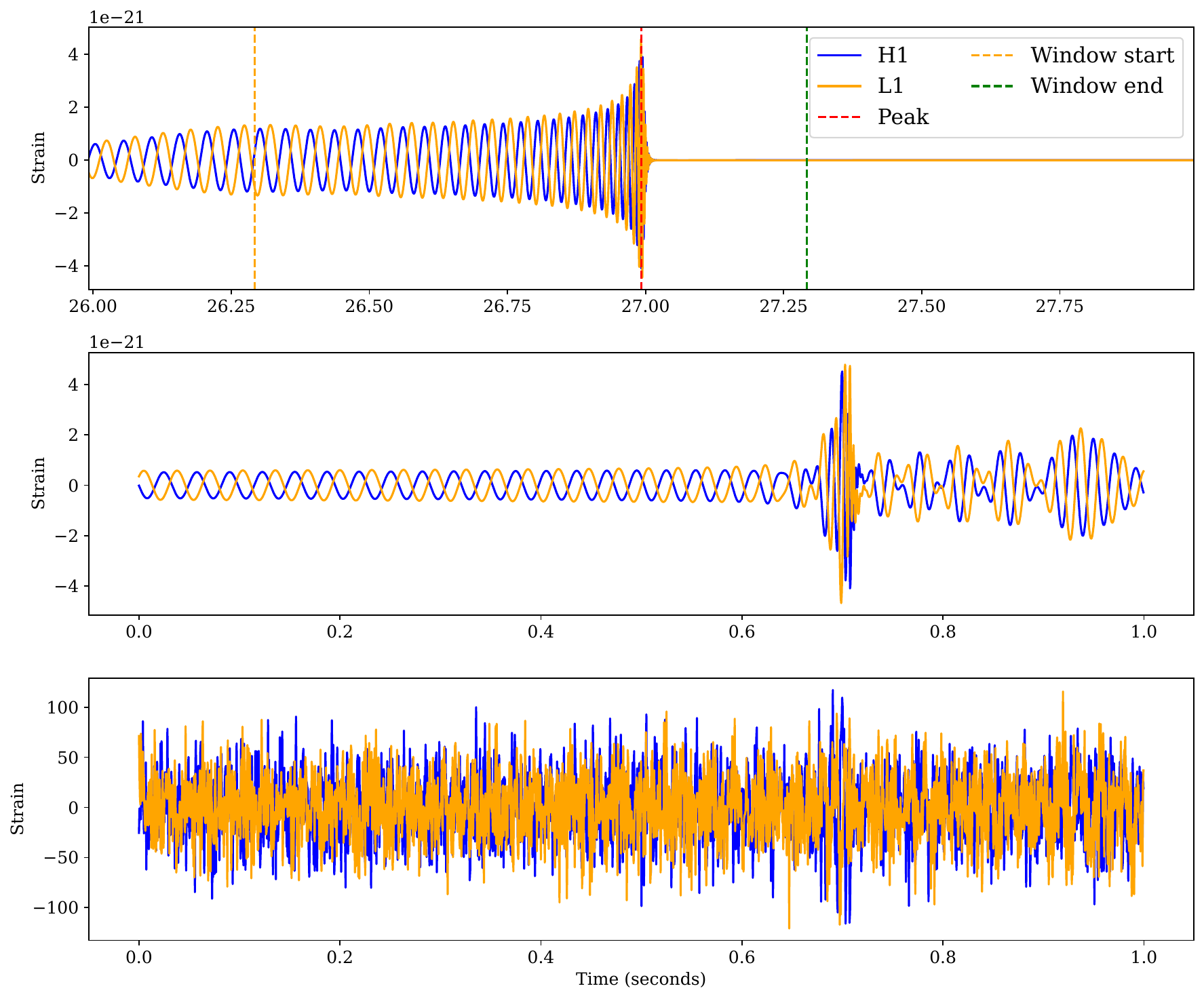}
\centering \caption{\label{fig1}Simulated GW signal. The horizontal axis represents time, while the vertical axis represents strain. Top figure: A raw GW signal is generated from randomly sampled BBH parameters. Middle figure: The GW signal within a 1-second time window is extracted from the original signal. Bottom figure: Colored noise is injected into the extracted signal segment, which is then whitened and normalized.}
\end{figure*}

This work employs a ppE description of the excitation phase as its physical core when simulating GW signals incorporating GR deviations. Specifically, we adopted an approach based on the TIGER framework \cite{Agathos_2014, Meidam_2018, PhysRevD.103.122002, Li_2012}. In the frequency domain, the phase can be expressed as
\begin{equation}
\Psi(f) = 2\pi ft_c - \phi_0 - \frac{\pi}{4} + \sum_{j=0}^{7} \left[ \psi_j + \psi_j^{(l)} \ln f \right] f^{(j-5)/3},
\end{equation}
where $t_c$ denotes the merging time, $\phi_0$ is the reference phase, and the coefficients $\psi_j$ and $\psi_j^{(l)}$ represent the PN phase terms in GR, expressed as functions of intrinsic binary parameters. TIGER modifies these GR coefficients to
\begin{equation}
\psi_i = \left( 1 + \delta\chi_i \right) \psi_i^{\text{GR}}.
\end{equation}
The signal injection process is shown in Fig.~\ref{fig1}. Before feeding data into the neural network, we perform a final normalization step: linearly normalizing the parameter vector from its original physical range to [-1, 1]. This helps stabilize the neural network training process. Additionally, for each whitened strain sequence $x$, we perform zero-mean and unit-variance normalization
\begin{equation}
\hat{x} = \frac{x - \mu_x}{\sigma_x},
\end{equation}
where $\mu_x$ and $\sigma_x$ are the mean and standard deviation of that specific strain sequence. This operation preserves the relative shape of the signal while accelerating model convergence. 

\subsection{Deep learning models}\label{sec2.2}

Deep learning specifically employs neural networks with multiple layers \cite{LeCun2015}, and can automatically learn hierarchies directly from raw data through its layered architecture \cite{Zeiler2014}. The traditional Bayesian theorem is
\begin{equation}
P(\theta \mid x) = \frac{P(x) P(x \mid \theta)}{P(\theta)}.
\end{equation}
Among these, $P(x\mid\theta)$ is the likelihood function. In complex scientific models, this likelihood function is difficult to compute analytically or involves extremely high computational costs. The core idea of NPE is to bypass direct modeling of the likelihood function and instead use neural networks to directly learn the mapping from observed data $x$ to the posterior distribution of parameters $P(\theta\mid x)$. Specifically, this involves learning a conditional probability distribution $Q_\phi(\theta \mid x)$ parameterized by the neural network weights $\phi$, that approximates the true posterior distribution $P(\theta \mid x)$ given by Bayes' theorem. The network is trained in a supervised manner using a large set of parameter-data sample pairs $(\theta, x)$ generated by numerical simulation. The training process aims to minimize the discrepancy between the approximate posterior $Q_\phi(\theta \mid x)$ and the true posterior. In practice, this is achieved by maximizing the conditional log-likelihood $\log Q_\phi(\theta \mid x)$ as a surrogate objective. Once trained, the network can perform rapid inference. For a new observational datum $x_{\text{obs}}$, feeding it into the network directly yields the full conditional distribution $Q_\phi(\theta \mid x_{\text{obs}})$ as an approximation to the true posterior $P(\theta \mid x_{\text{obs}})$. 

\begin{figure*}[!htp]
\centering
\includegraphics[width=1.0\textwidth]{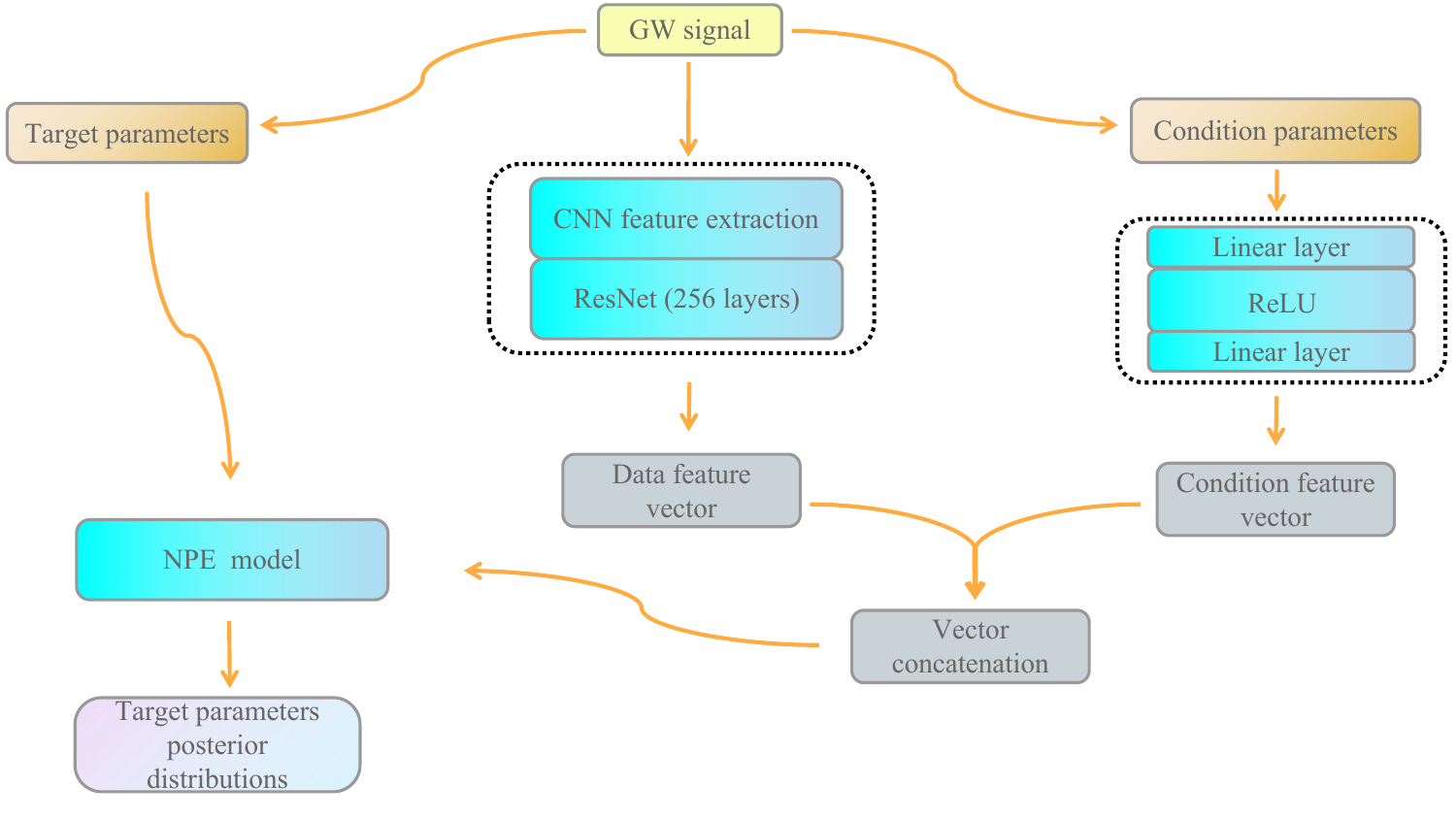}
\centering \caption{\label{fig2}Model architecture diagram. Both input data and parameters undergo normalization preprocessing. The dashed box on the left represents the hybrid residual embedding network, while the dashed box on the right denotes the conditional embedding network.}

\end{figure*}

Traditional NPE framework employs a single high-dimensional flow model to estimate the full posterior $P(\theta\mid x)$, facing the curse of dimensionality in a high-dimensional parameter space. We propose a conditional independence decomposition strategy on top of the NPE paradigm, approximating the joint posterior as
\begin{equation}
P(\theta | x) \approx P(\theta_{\text{basic}} | x) \prod_{i=7}^{15} P(\delta\chi_i | x, \theta_{i}^{\text{*}}).
\end{equation}
Here, $\theta_{i}^{\text{*}}$ denotes the parameters excluding $\delta\chi_i$. This approximation transforms the problem into 9 condition estimation tasks that can be optimized in parallel. It simplifies the high-dimensional joint estimation problem into a series of sub-tasks conditioned on partial parameters and their known physical relationships, significantly reducing model complexity and training difficulty. Second, it explicitly leverages the physical relationships between parameters as conditional information, thereby enhancing estimation accuracy. The architecture of the NPE model for each ppE parameter is shown in Fig.~\ref{fig2}. In our analysis, we allowed the ppE parameters to vary freely in turn, with the remaining parameters fixed to their GR values of zero. This approach is fully robust for detecting deviations present at multiple PN orders \cite{Sampson:2013lpa}.

To jointly capture local temporal features and global dependency patterns, we designed a hybrid embedding network that integrates a CNN with a deep ResNet. The preprocessed strain data is first processed by a three-layer one-dimensional CNN for downsampling and feature extraction, outputting a feature tensor. This tensor is then flattened and fed into a custom deep ResNet \cite{he2015deepresiduallearningimage}. This ResNet is composed of a stack of 128 residual blocks. Each block contains two fully-connected layers with 512 neurons, followed by batch normalization and dropout layers, and employs identity mapping via skip connections. The final output is a data embedding vector, which is used for subsequent parameter posterior estimation. This design provides the network with a strong capacity for nonlinear mapping and feature abstraction, allowing it to capture subtle correlations between GW signals and the target parameters. Moreover, Such deep architectures enable the learning of complex mappings directly from noisy data \cite{Nousi_2023}. Then we introduce a conditional embedding network, which refers to a module that encodes ancillary physical parameters to guide the inference. GW parameters exhibit strong physical coupling, with mass and spin jointly determining the waveform evolution. Neglecting such correlations when estimating deviations from GR would introduce significant bias. To address this, we explicitly condition the inference on the remaining physical parameters, denoted as $\theta_{i}^{\text{*}}$. These conditional parameters are encoded via a multi-layer perceptron (MLP) to generate conditional embedding vectors. The MLP comprises two fully connected layers. The first layer projects the input parameters into a high-dimensional latent space, employing a ReLU \cite{10.5555/3104322.3104425} activation function to introduce nonlinearity. The second layer compresses this representation into a compact conditional embedding. {The MLP is designed to encode the strong nonlinear degeneracies between the chirp mass and the ppE parameters ($\delta\chi_0$ through $\delta\chi_2$), which is particularly critical for low-SNR signals. This correlation information is propagated to each factorized ppE model, thereby helping to prevent artificial decoupling of the parameters.}

During training, the conditional embedding network and the hybrid residual network operate in parallel. The output feature vectors from both networks are concatenated at a fusion layer and then passed to the NPE model. The NPE model is built by stacking 10 affine coupling layers. Each coupling layer incorporates a sub-network with 5 hidden layers of 512 neurons each. The model directly outputs the conditional posterior distribution of the target deviation parameter, $P(\delta \chi_i \mid x, \theta_{i}^{\text{*}})$. This architecture allows the posterior estimate to explicitly incorporate constraints from known physical parameters, effectively mitigating issues such as mode collapse and posterior degeneracy in high-dimensional parameter spaces. Furthermore, the conditional embedding mechanism enables a factored estimation strategy. For each of the ppE parameters $\delta\chi_0$ to $\delta\chi_7$, the network performs an independent estimation conditioned on all other parameters. This approach significantly reduces the modeling complexity for each estimator while preserving consistent physical constraints across parameters via a shared conditional encoder. The architecture accommodates flexible conditional input during inference. When values for some conditional parameters are unknown, plausible estimates can be generated by sampling from prior distributions or through iterative optimization, offering a scalable solution for parameter inference with real observational data.

\subsection{Strategy}\label{sec2.3}
\begin{figure*}[!htp]
\centering
\includegraphics[width=1.0\textwidth]{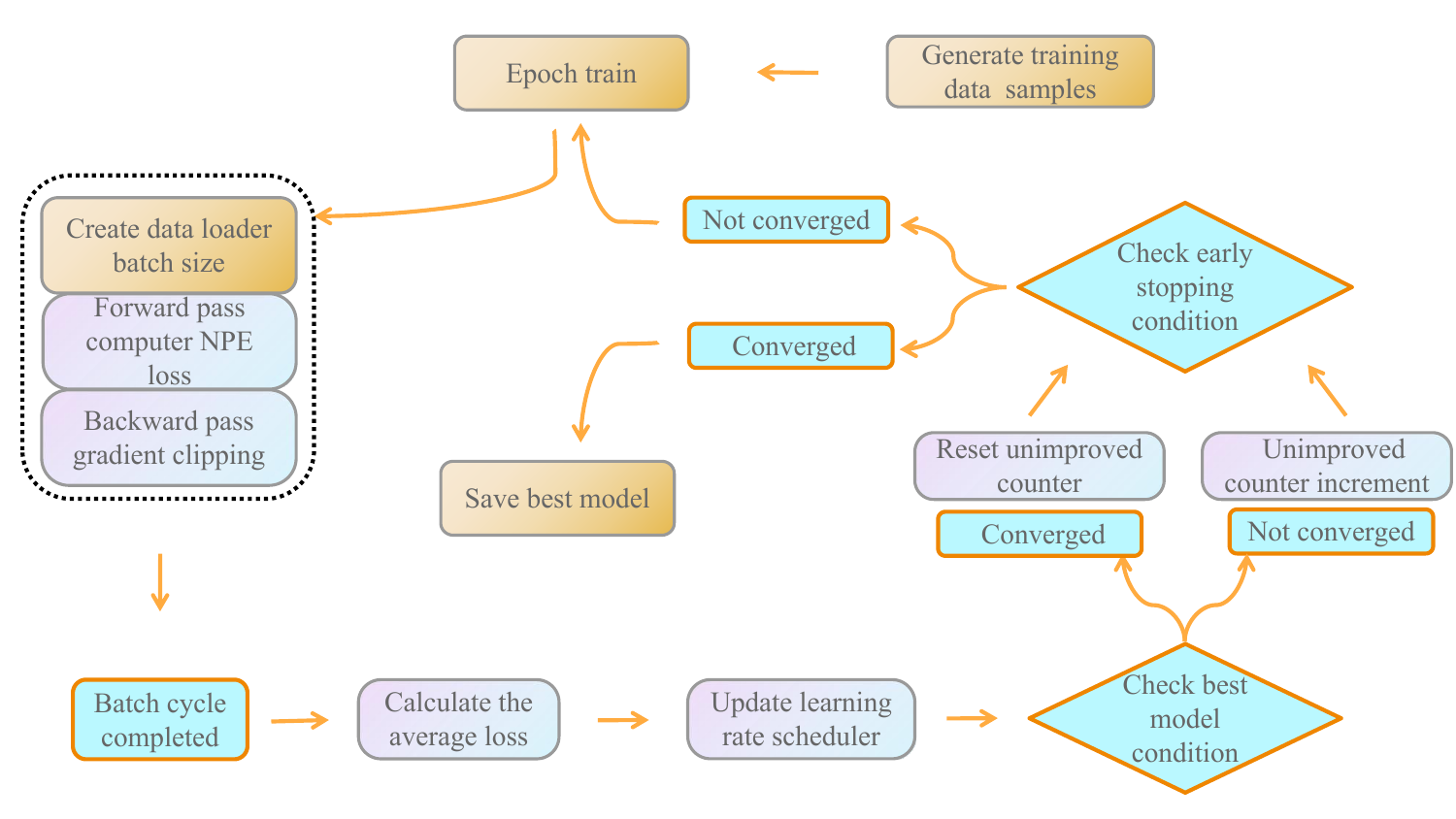}
\centering \caption{\label{fig3}Training strategy principles. The dashed box on the left represents the inner loop for batch data.}
\end{figure*}

We employ the negative log-likelihood loss as the objective function for training each estimator, consistent with the theoretical foundation of NPE \cite{papamakarios2018fastepsilonfreeinferencesimulation,2021Normalizing}. This loss for the i-th estimator is defined as
\begin{equation}
L(\phi_i) = -\frac{1}{N} \sum_{k=1}^{N} \log Q_{\phi_i} \left((\delta \chi_i^{(k)}) \mid x^{(k)}, \theta_{i}^{\text{*}^{(k)}} \right).
\end{equation}
Each model is trained independently using the AdamW optimizer \cite{loshchilov2019decoupledweightdecayregularization} with an initial learning rate \(1 \times 10^{-3}\), a weight decay coefficient of \(1 \times 10^{-4}\). Training was performed with a batch size of 2048 for a maximum of 200 epochs, incorporating an early stopping criterion to improve efficiency. To mitigate overfitting, in addition to the weight decay regularization applied by the optimizer. Dropout with a rate of 0.1 was applied to all fully-connected layers within the network. Throughout the training process, the model's loss was continuously monitored on a separate validation set. The early stopping mechanism was triggered if the validation loss failed to decrease for 15 consecutive epochs. Upon triggering, the training process was halted, and the model parameters were reverted to the snapshot corresponding to the epoch with the lowest validation loss, thereby ensuring optimal generalization performance. This training and validation procedure is illustrated in Fig.~\ref{fig3}.

\begin{table*}[t]
\caption{Posterior distribution statistics for ppE parameters. Values are presented as median with upper and lower 90\% credible intervals.}
\label{tab1}
\centering
\setlength\tabcolsep{18pt}
\renewcommand{\arraystretch}{1.5}
\begin{tabular}{lcc}
\hline \hline  
Parameter & NPE (90\% confidence interval) & MCMC (90\% confidence interval)\\
\hline  
$\delta\chi_0$ & $0.1941^{+0.8423}_{-0.0531}$ & $-0.0279^{+0.7888}_{-0.6456}$ \\
$\delta\chi_1$ & $-0.8210^{+0.4033}_{-0.9842}$ & $-0.5990^{+0.1837}_{-0.9792}$ \\
$\delta\chi_2$ & $0.6948^{+0.9820}_{-0.1516}$ & $0.3621^{+0.9435}_{-0.7060}$ \\
$\delta\chi_3$ & $0.2219^{+0.9189}_{-0.7929}$ & $0.0463^{+0.7825}_{-0.6874}$ \\
$\delta\chi_4$ & $0.0569^{+0.8917}_{-0.9071}$ & $-0.0128^{+0.8698}_{-0.8718}$ \\
$\delta\chi_{5l}$ & $-0.0300^{+0.8951}_{-0.9083}$ & $0.1966^{+0.9027}_{-0.8350}$ \\
$\delta\chi_6$ & $0.0667^{+0.9016}_{-0.8955}$ & $0.4267^{+0.9509}_{-0.5219}$ \\
$\delta\chi_{6l}$ & $-0.0219^{+0.8957}_{-0.9127}$ & $-0.0285^{+0.8490}_{-0.8752}$ \\
$\delta\chi_7$ & $0.0039^{+0.9183}_{-0.9138}$ & $-0.0321^{+0.8579}_{-0.8769}$ \\
\hline \hline
\end{tabular}

\end{table*}

The implementation in this work was developed with Python 3.10.18 and the PyTorch framework (version 2.5.1) \cite{NEURIPS2019_9015}. All model training and evaluation were performed on a computational server equipped with an NVIDIA RTX A6000 GPU, utilizing CUDA 12.4 for acceleration. Training each ppE parameter estimator required approximately 8 to 13 hours, depending on the convergence speed. After training, inference time for parameter estimation on a single event was less than 0.4 seconds. Compared to traditional Bayesian methods that typically require hours, this efficiency lays the foundation for real-time analysis of GW parameters. To ensure frequency rigor of the posterior estimates, we implemented systematic Monte Carlo coverage tests \cite{hermans2022trustcrisissimulationbasedinference}. For each ppE parameter, we computed its empirical coverage based on a test set containing 1000 independent simulated events. If the posterior distribution is accurate, 
\begin{equation}
u_i = \int_{-\infty}^{\delta \chi_{i, \text{true}}} P(\delta \chi_i \mid x, \theta_{i}^{\text{*}}) \, \textcolor{blue}{\mathrm{d}}(\delta \chi_i )
\end{equation}
should be uniformly distributed over the interval [0, 1]. Here, $u_i$ represents the sequence of probability integral transform (PIT) values. We employ the KS \cite{Lopes2011} test to quantify the deviation between the empirical distribution and the theoretical uniform distribution, and evaluate the performance by computing the empirical coverage at multiple confidence levels. This test is repeated for each independent model to ensure no systematic bias.

\section{RESULTS AND DISCUSSION}\label{sec3}

\subsection{Posterior distribution quality}\label{sec3.1}
\begin{figure*}[!htp]
\centering
\includegraphics[width=1.0\textwidth]{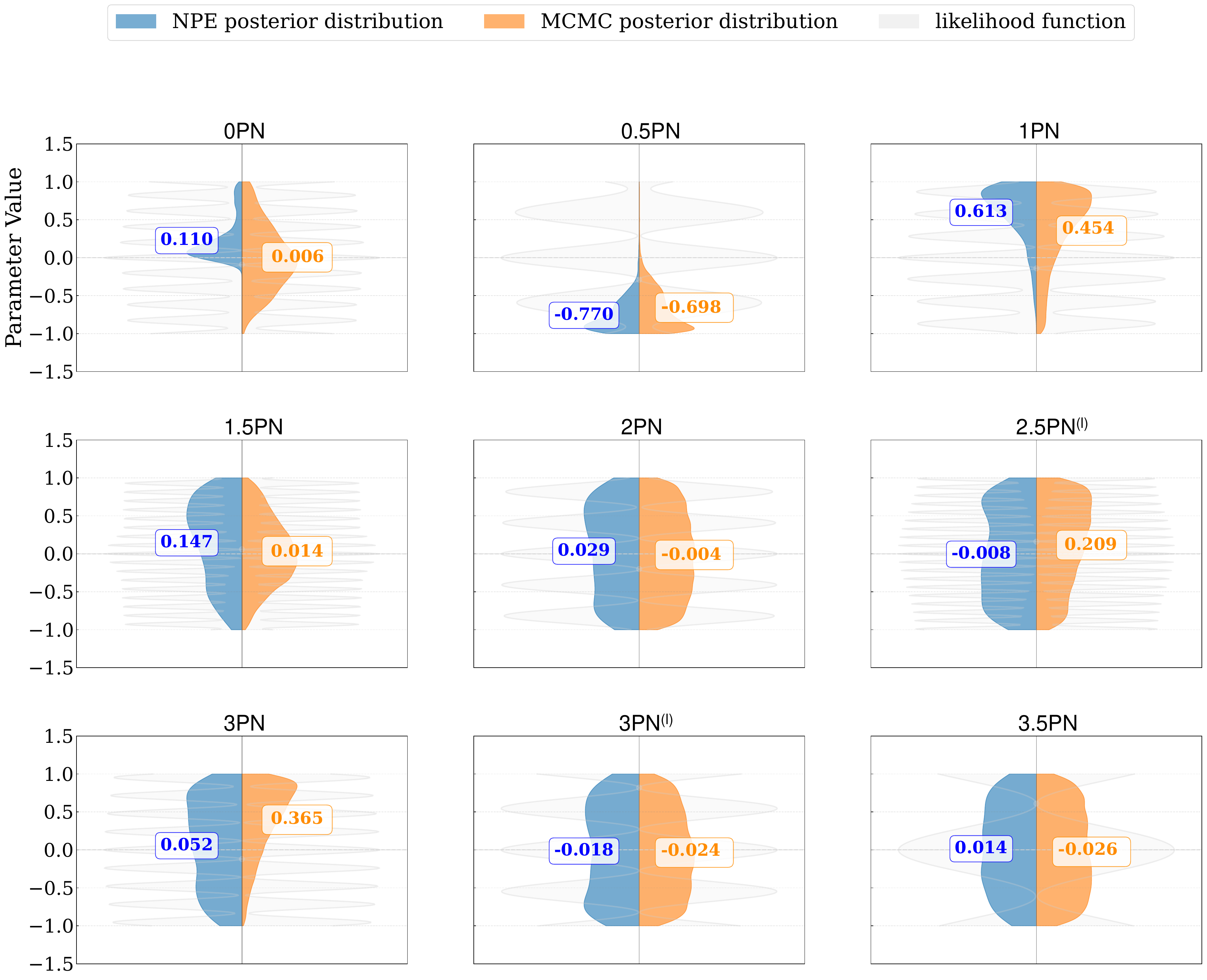}
\centering \caption{\label{fig4}Posterior distributions of ppE parameters. The left side of the violin plot shows the results from the deep learning based NPE method, while the right side shows the results from the MCMC method based on the {\tt Dynesty} nested sampler. The values labeled in the violin plot represent the mean of the posterior distribution obtained by each method. The gray background shows the likelihood function distribution of the parameter. The labels 0PN to 3.5PN on the diagram correspond to $\delta\chi_0$ to $\delta\chi_7$.}
\end{figure*}

We compare the posterior distribution estimates for the 9 ppE parameters from the MCMC and NPE methods, with the results presented in Fig.~\ref{fig4}. This figure employs a split violin plot format to visually illustrate the differences and similarities between the two methods in estimating the posterior distributions of the target parameters and innovatively employs the symmetric likelihood curve as the background. 

In the MCMC method, the violin plots of the posterior distributions exhibit relatively symmetric shapes, with the means of all parameters close to their true values, indicating that this method can accurately estimate the posterior distributions of parameters. 
We perform Bayesian inference using the {\tt Bilby} \cite{Ashton_2019} software package with the {\tt Dynesty} nested sampler. The prior ranges of the parameters are the same as those used in the NPE method. The nested sampling is configured with 4096 live points, uses the random walk sampling method, and has a termination criterion with dlogz set to 0.1.
However, this method incurs high computational costs, particularly in high-dimensional parameter spaces. 

In contrast, the violin plots from the NPE framework exhibit similar shapes to MCMC results, but the distributions for some parameters appear slightly narrower, indicating that the posterior distributions estimated by NPE possess lower uncertainty. For parameters $\delta \chi_3$ to $\delta \chi_7$, the violin plot widths from MCMC and NPE are largely consistent, indicating comparable uncertainty estimates between the two methods. For the $\delta \chi_0$ to $\delta \chi_2$ parameters, the NPE method exhibits a slightly narrower distribution. To further compare the inference results of the two methods, Table~\ref{tab1} lists the 90\% confidence intervals for parameters $\delta \chi_0$ to $\delta \chi_7$. The core commonality lies in the fact that the confidence intervals provided by both methods encompass the true parameter value. This directly confirms that, under the testing scenario defined in this work, our NPE model and the traditional MCMC method used as a benchmark yield the identical core physical conclusion that no significant deviation from GR has been detected.

Specifically, for parameters $\delta\chi_0$ to $\delta\chi_2$, the posterior distribution intervals provided by NPE are significantly narrower than those obtained via MCMC. 
The BBH waveforms consist of multiple harmonic components superimposed. This structure can induce several secondary peaks in addition to a primary peak within the parameter space. The primary peak indicates that all harmonics are well-matched, while secondary peaks suggest that only some harmonics are well-fitted \cite{Zou:2024osb}. These secondary peaks are numerous, widely distributed in parameter space, and often comparable in amplitude to the global primary peak. Their presence makes signal detection relatively straightforward, as any prominent peak can serve as evidence that the data contain a signal. However, these peaks also increase the estimation error for signal parameters by introducing parameter degeneracy \cite{Zou:2024jqv}. A MCMC sampler explores the parameter space via random walks from an initial point. If multiple posterior peaks are separated by regions of low probability, the acceptance probability for transitions between peaks becomes exceedingly low. Consequently, the chain can become trapped in the neighborhood of the first peak it encounters, preventing the effective exploration of other peaks with comparable posterior probability and leading the sampling to converge to a local optimum. The resulting posterior distribution will misrepresent that local morphology as a global characteristic. This occurs because the chain fails to explore the remaining probability mass, leading to a severe overestimation of the actual probability density within that region. 

However, the NPE framework is based on amortized posterior estimation. It uses a normalizing flow to learn a direct mapping from the data space to the parameter posterior distribution. Modern normalizing flow architectures inherently possess the capacity to represent complex, multimodal distributions. During training, the network processes a large set of paired samples of parameters and corresponding simulated data, drawn from the full prior distribution. Through this process, it learns the association between data features and the multiple high-probability regions in parameter space. Provided that the training simulation adequately covers the multimodal structure of the parameter space, the neural network can simultaneously learn the location and shape of all major modes. Once trained, the NPE model can, for a given observed data instance, directly generate a complete posterior distribution. This distribution can naturally contain several distinct, separated modes, to each of which it assigns a relatively accurate probabilistic weight.

\subsection{Overall performance of ppE parameters estimation and frequency calibration}\label{sec3.2}
The KS test \cite{Lopes2011} is a non-parametric statistical method used to assess the consistency between an empirical sample distribution and a reference theoretical distribution. In GW parameter estimation, this method is frequently employed to evaluate whether the inferred posterior distribution exhibits good statistical calibration \cite{Sidery:2013zua,Biwer:2018osg,Thrane_Talbot_2020}. The KS test operates by computing and comparing the cumulative distribution functions of the two distributions. The maximum absolute distance between them is defined as
\begin{equation}
D_{n} = \max \mid F_{n}(x)-F(x)\mid,
\end{equation}
where $D_n$ is the test statistic, $F_n(x)$ is the empirical cumulative distribution function and $F(x)$ is the theoretical cumulative distribution function. The $p$-value, derived from $D_n$, quantifies the probability that the observed discrepancy between the distributions is due to random chance. A significance level threshold of 0.05 is typically adopted. If $p < 0.05$, the null hypothesis of identical distributions is rejected, indicating a statistically significant difference.

\begin{figure}[!htp]
\centering
\includegraphics[width=7.6cm, height=8.5cm]{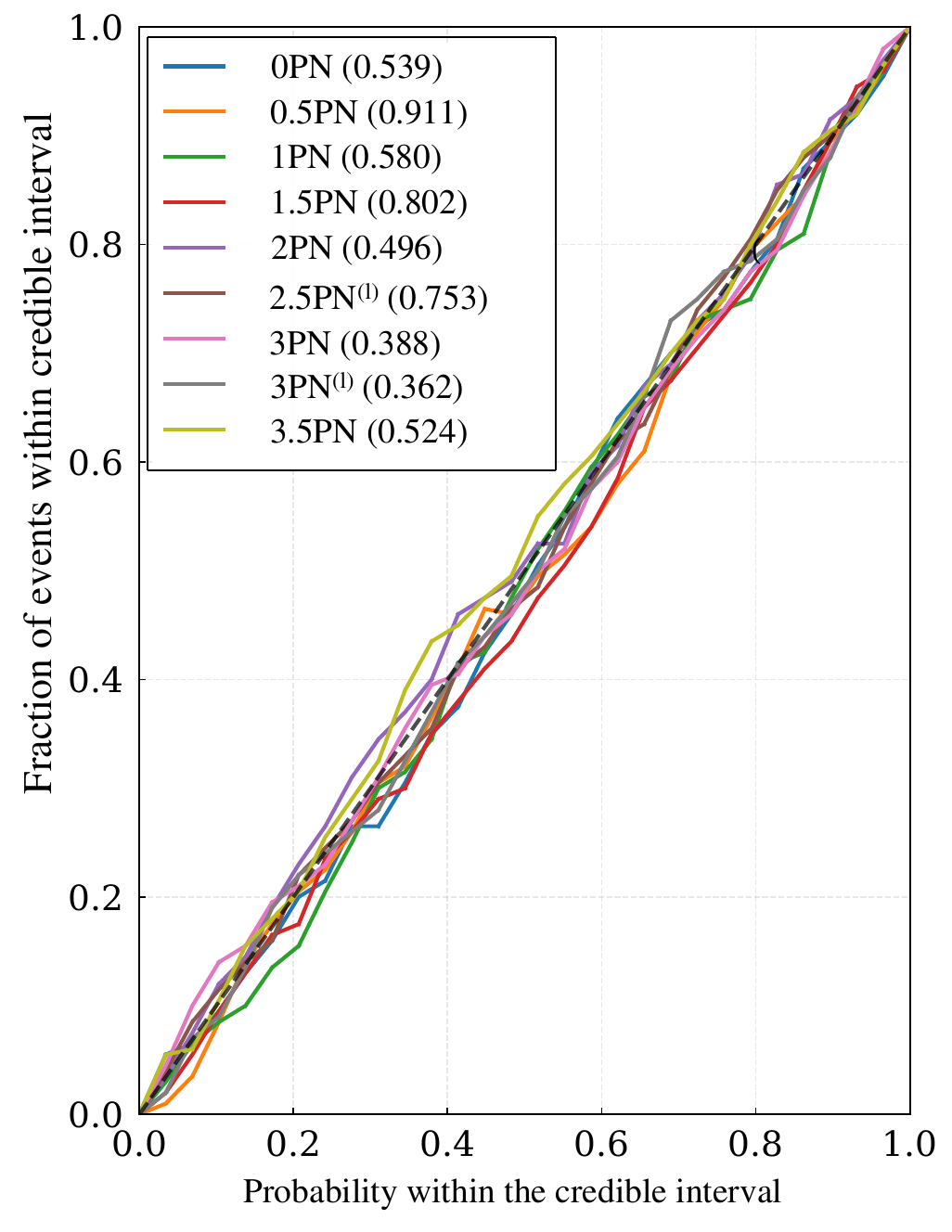}
\centering \caption{\label{fig5}Coverage probability curves of ppE parameters from 1000 injections. The labels 0PN to 3.5PN on the diagram correspond to $\delta\chi_0$ to $\delta\chi_7$.}
\end{figure}

In this work, we utilize the KS test to evaluate the calibration quality of the NPE model. Specifically, we compare two distributions: the first is the theoretical uniform distribution $U(0,1)$, and the second is a sequence of PIT values computed from the estimated posterior. The latter consists of the probabilities that the true parameter values fall within the estimated credible intervals across multiple simulations. Our null hypothesis posits that the two sets of samples are drawn from the same distribution, i.e., the model posterior is perfectly calibrated. By computing the KS statistic and its corresponding $p$-value, and comparing it against the 0.05 significance level, we determine whether to reject the null hypothesis, thereby providing a quantitative diagnostic of the statistical reliability of the posterior estimates.
To validate the effectiveness of the NPE model for inference on individual parameters, we conducted a calibration test based on simulated data. The test comprised 1000 independent simulated event injections. We computed the frequency at which true parameter values were contained within the corresponding posterior credible intervals across multiple confidence levels. For each event, we calculated the percentile corresponding to the true parameter value within its estimated posterior cumulative distribution function. Figure~\ref{fig5} displays the joint empirical cumulative distribution function of these percentiles across all events. Under conditions of perfect calibration, this distribution should follow a uniform distribution between 0 and 1, which corresponds to the diagonal line in the figure. The deviation of the estimated curve from this diagonal directly reflects the bias in the posterior estimates; a closer alignment indicates better calibration performance of the parameter inference. The $p$-values obtained from the KS test, annotated in the figure legend, provide a quantitative assessment of the agreement between the empirical distribution and the theoretical uniform distribution. 

The results demonstrate that the NPE model exhibits well-calibrated performance. Specifically, parameters $\delta\chi_1$, $\delta\chi_3$, $\delta\chi_{5l}$ show particularly outstanding calibration performance, with their coverage probability curves nearly coinciding with the ideal line.

\begin{table}
\caption{Computational performance comparison.}
\label{tab2}
\centering
\setlength\tabcolsep{10pt}
\renewcommand{\arraystretch}{1.5}
\begin{tabular}{lcc}
\hline \hline
&MCMC & NPE \\
&(Run time/second) &(Run time/second)\\
\hline 
 $\delta\chi_0 $&27449& 0.31\\
 $\delta\chi_1 $& 26831&0.39\\
 $\delta\chi_2 $& 33846& 0.34\\
 $\delta\chi_3 $& 28501& 0.35\\
 $\delta\chi_4 $& 22799& 0.38\\
 $\delta\chi_{5l}$& 36205& 0.38\\
 $\delta\chi_6 $& 44818& 0.21\\
 $\delta\chi_{6l} $& 20943& 0.35\\
 $\delta\chi_7 $& 29021& 0.36\\ 
 \hline \hline
\end{tabular}

\end{table}

\subsection{Computational performance comparison}\label{sec3.3}
As shown in Table~\ref{tab2}, the NPE method is faster than traditional MCMC by a factor of approximately \(9 \times 10^4\). The primary computational cost is concentrated in an offline training phase, which involves approximately {8 to 13} hours of neural network optimization. This training cost constitutes a fixed, upfront investment that can be amortized over subsequent inferences for any number of observed events. Once trained, performing complete posterior sampling and parameter estimation for a single GW event requires only about 400 milliseconds. This efficiency gain makes it feasible to obtain preliminary constraints on the ppE framework parameters within a very short time after a GW signal is detected, thereby providing crucial inference results to inform subsequent observational decisions.

\begin{figure*}[!htp]
\centering
\includegraphics[width=1.0\textwidth]{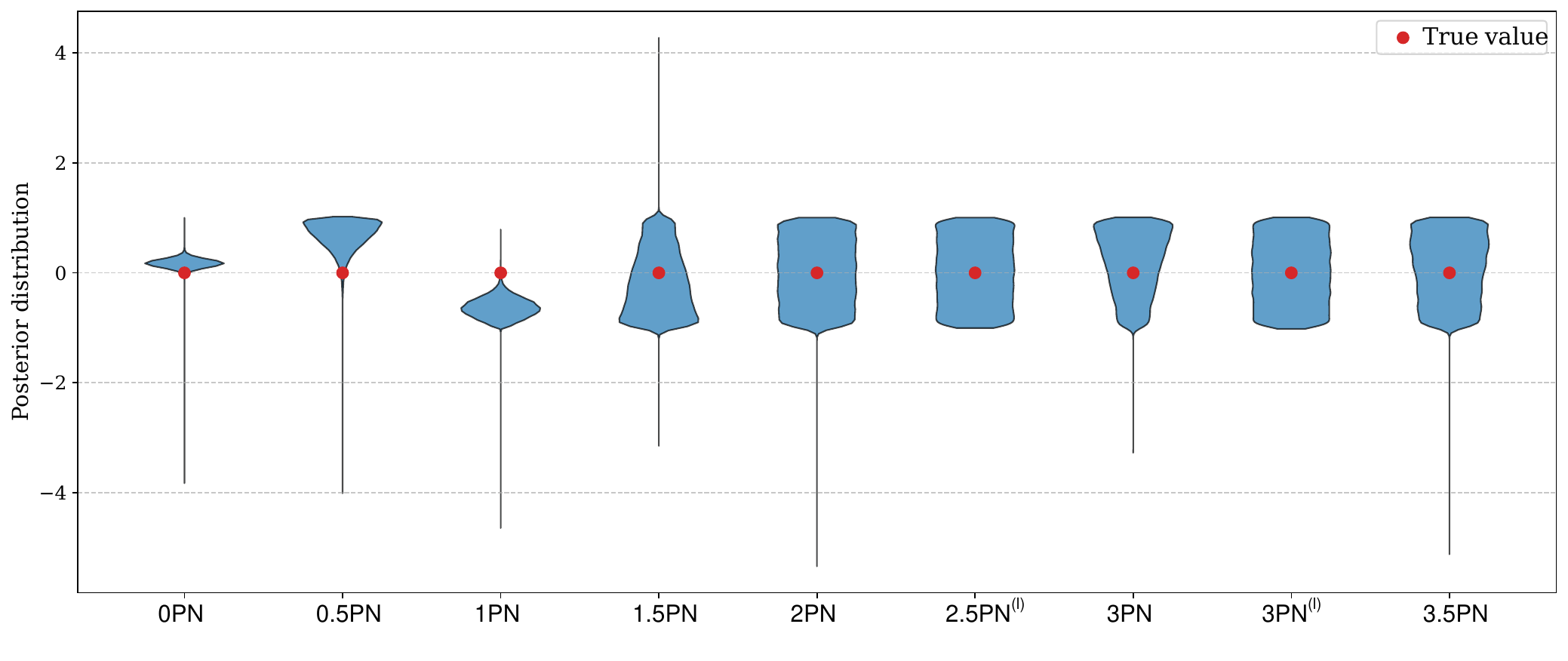}
\centering \caption{\label{fig6}The posterior distribution of the ppE parameters for the GW150914 event. The labels 0PN to 3.5PN on the diagram correspond to $\delta\chi_0$ to $\delta\chi_7$.}
\end{figure*}

\begin{figure*}[!htp]
\centering
\includegraphics[width=1.0\textwidth]{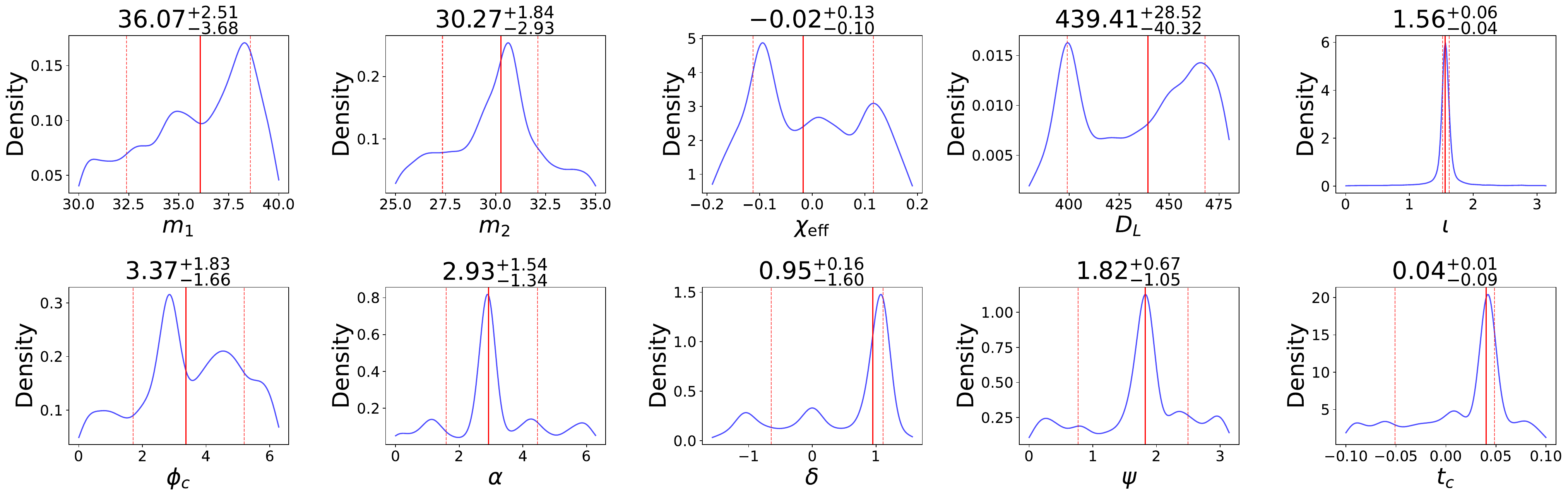}
\centering \caption{\label{fig7}Status of the restoration of GW150914 physical source parameters. The solid red line indicates the median, and the interval between the two dashed red lines corresponds to the 90\% credible interval. {The posterior median and the 90\% credible interval for each parameter are indicated at the top of each subplot.}}
\end{figure*}

\subsection{Analysis of event GW150914 }\label{sec3.4}

To evaluate the performance of our factorized NPE framework on real observational data, we apply it to GW150914, the first directly detected GW event. We apply the same preprocessing pipeline used for the training set to the real data \cite{LIGOScientific:2018mvr}. The processed strain data are then input into the trained networks. For each ppE deviation parameter, the network directly outputs its posterior distribution. In addition to the ppE deviation parameters, accurately recovering the standard BBH parameters is a prerequisite for any test of GR. Therefore, we evaluate the performance of our framework in inferring these fundamental physical parameters for GW150914. The results are shown in Fig.~\ref{fig6}, Fig.~\ref{fig7}.

It is worth noting that our training set is based on simulated Gaussian noise, whereas the real noise of GW150914 is non‑stationary. Ignoring this difference in noise characteristics may not introduce a systematic bias in parameter estimation. However, it may distort the shape of the posterior distribution, thereby affecting the accuracy of uncertainty estimates \cite{Edy:2021par}. {For the parameters $\delta\chi_{3}$ to $\delta\chi_{7}$, the marginal posterior distributions exhibit bilaterally stretched tails, overall widened credible intervals, and the emergence of multimodal structures. For the parameters $m_1$, $m_2$, $\chi_{\text{eff}}$, $D_L$, non-stationary noise moderately narrows the upper and lower credible boundaries. These effects primarily influence the precision of uncertainty quantification and the fine morphology of the posterior, rather than the central estimates. A more rigorous investigation of the impact of non-Gaussian and non-stationary noise on our framework is planned for future work.}  Therefore, the posterior distribution obtained using our NPE framework is generally consistent with existing inferences regarding this event \cite{LIGOScientific:2018mvr,LIGOScientific:2016dsl}. As shown in Table ~\ref{tab3}, this finding confirms that our framework is capable of deriving physical conclusions from real data that are consistent with standard analyses.

\begin{table*}[!htp]

\caption{Posterior distribution statistics for BBH parameters. Values are presented as median with upper and lower 90\% credible intervals. The table lists only the parameters most representative of the GR test.}
\label{tab3}
\centering
\setlength\tabcolsep{18pt}
\renewcommand{\arraystretch}{1.5}
\begin{tabular}{lcc}
\hline \hline  
Parameter & NPE (90\% confidence interval) & MCMC (90\% confidence interval)\\
\hline  
$m_1$ & $36.07^{+2.51}_{-3.68}$ & $35.6^{+4.7}_{-3.1}$ \\
$m_2$ & $30.27^{+1.84}_{-2.93}$ & $30.6^{+3.0}_{-4.4}$ \\
$\chi_{\text{eff}}$ & $-0.02^{+0.13}_{-0.10}$ & $-0.01^{+0.12}_{-0.13}$ \\
$D_L$ & $439.41^{+28.52}_{-40.32}$ & $440^{+150}_{-170}$ \\
\hline \hline
\end{tabular}

\end{table*}

\section{CONCLUSION}\label{sec4}

For the problem of testing GR with BBH merger signals, this work proposes a parameter-independent NPE framework. This framework enables efficient and rigorous inference of the 9 ppE parameters in such signals. We employ a hybrid deep ResNet to extract data features. By introducing a training strategy based on conditional embedding and parameter separation, we decompose an independent normalizing flow model for each ppE parameter, significantly reducing the complexity of model training.

The results indicate that for all 9 ppE parameters, the posterior estimates from our NPE framework are consistent with those from the traditional MCMC nested sampling benchmark. The KS test verifies the uniformity of the posterior cumulative probabilities, indicating that the framework's inferences satisfy rigorous frequentist statistical requirements. In terms of computational efficiency, our framework requires only approximately 0.4 seconds to perform complete posterior sampling for a single event, whereas the traditional method based on MCMC nested sampling requires over 10 hours, achieving an acceleration factor of approximately $9 \times 10^4$. In addition, we conducted tests using GW150914, and the posterior distribution obtained from the NPE framework agreed well with the official results.

With its high computational performance and rigorously validated statistical reliability, the method proposed in this work provides a scalable and verifiable technical solution for testing GR, designed for next-generation observational facilities such as the Einstein Telescope \cite{Maggiore_2020,Punturo2010TheET,abac2025scienceeinsteintelescope} and space-based GW detectors (e.g., Taiji \cite{Luo2020TheTP} and TianQin \cite{Hu_2018}). This approach not only holds the potential to enhance the sensitivity and efficiency of searching for signatures of physics beyond GR in GW signals but also demonstrates the potential of artificial intelligence-driven inference paradigms in advancing the forefront of GW astrophysics.

\begin{acknowledgments}
We are grateful to Ji-Yu Song for insightful discussions. This research has made use of data or software obtained from the Gravitational Wave Open Science Center (gwosc.org), a service of LIGO Laboratory, the LIGO Scientific Collaboration, the Virgo Collaboration, and KAGRA. This work was supported by the National Natural Science Foundation of China (Grants Nos. 12473001, 12503001, and 12575049), the National SKA Program of China (Grants Nos. 2022SKA0110200 and 2022SKA0110203), the China Manned Space Program (Grant No.CMS-CSST-2025-A02), the 111 Project (Grant No. B16009), and the China Scholarship Council.
\end{acknowledgments}

\bibliography{reference}

\end{document}